\documentstyle[12pt]{article}
\newcommand \be {\begin{equation}}
\newcommand \bea {\begin{eqnarray}}
\newcommand \ee {\end{equation}}
\newcommand \eea {\end{eqnarray}}

\newcommand{\bit}{\begin{itemize}}
\newcommand{\eit}{\end{itemize}}

\begin{document}
\topskip 2cm

\begin{titlepage}

\begin{center}
{\large\bf A note on Seiberg-Witten central charge.} \\
\vspace{2.5cm}
{\large Alfredo Iorio} \\
\vspace{.5cm}
{\sl  School of Mathematics, Trinity College}\\
{\sl College Green, Dublin 2, Ireland} \\
{\sl and} \\
{\sl School of Theoretical Physics,}
{\sl Dublin Institute for Advanced Studies}\\
{\sl 10 Burlington Road, Dublin 4, Ireland} \\
\vspace {.2cm}
{\sl e-mail: iorio@maths.tcd.ie} \\
\vspace{.5cm}

\vspace{2.5cm}

\begin{abstract}
\noindent
The central charge for the Seiberg-Witten low-energy effective Action
is computed using Noether supercharges. A reliable method to construct 
supersymmetric Noether currents is presented.   
\end{abstract}
\end{center}

\vspace{2.5cm}

\end{titlepage}

\newpage

\noindent
The electromagnetic duality of Seiberg-Witten \cite{sw} relies heavily on 
the BPS mass-formula

\be\label{Z}
M = \vert Z \vert \quad {\rm where} \quad  
Z = (n_e a + n_m a_D)
\ee
is the topological central charge of the N=2 
supersymmetry\footnote{Duality means  
$\left( \begin{array}{c} a_D \\ a \end{array} \right)
\rightarrow D \left( \begin{array}{c} a_D \\ a \end{array} \right)$ and
$( n_m \; n_e ) \rightarrow ( n_m \; n_e ) 
D^{-1}$ where $D \in SL(2, Z)$.}.  
In Ref. \cite{wo} the classical $Z$
was computed and it was 
conjectured that if the fields belong to the {\it small} representation
of the N=2 supersymmetry they should saturate the BPS mass-formula,
giving $M=\vert Z \vert$, even at the quantum level.
But so far the only 
direct evidence of this is a BPS-type computation \cite{roc} 
of the minimum of the 
Hamiltonian, neglecting fermionic fields\footnote{The kind of 
result we are considering
also seems to follow from a geometric 
analysis of the N=2 vector multiplet in Ref.
\cite{van}, where, however, 
the authors' aim there is completely different, the fermionic contribution 
is not present and there is no mention of Noether charges.}. 
As stressed by Seiberg and Witten, another way to find the modification to the 
classical BPS formula is to compute the central charge $Z$ from the low-energy
U(1) effective Action\footnote{Despite the fact that the 
{\it massive} SU(2) gauge theory 
does not have 
the exact duality symmetry
(due to the presence \cite{bil}
of the scalar-vector-superfield coupling $\phi^\dagger e^V$), 
one would also like to calculate the $Z$ for this theory, 
both to check that the U(1) argument 
given above is correct, and that the freedom to make a linear shift
${\cal F}(A) \rightarrow {\cal F}(A) + c A$
which appears in the U(1) theory is removed at the full SU(2) level.
The SU(2) case will be treated in a later paper\cite{ior2}} 
because only the {\it massless} (and neutral)
degrees of freedom contribute to $Z$.

\noindent
A way to compute the central charge $Z$ is from the  
commutation relations of the supersymmetric 
Noether charges. The purpose of this paper is to present 
the result of such a computation for the low-energy U(1) 
part of the effective Action. The computation is by no means trivial but 
at the end the formula (1) is exactly what is obtained. Apart from the result 
itself, it is illuminating to see precisely how the centre 
remains {\it formally insensitive} to the quantum corrections. 

\noindent
As for any space-time symmetry, 
the Noether procedure for supersymmetry needs to be handled with 
care because of the variation $\delta {\cal L}=\partial_\mu V^\mu$ 
of the Lagrangian density and we present below a reliable 
way to write supercurrents taking this term carefully into account.  
We also show how to commute the charges, paying due attention
to the way they are expressed in terms of fields and momenta. 

\noindent
In this letter we shall give only a brief account of the   
computation with special emphasis on the 
application of the Noether technique to Seiberg-Witten centre, 
leaving to later work a detailed discussion \cite{ior2}.
One of the most important results of the detailed discussion is that
the general form of a supersymmetric Noether current is 

\be
J^\mu = \Pi_i^\mu \delta \Phi^i \; - \; V^\mu
\ee
where the $\Phi_i$'s are the fields in the Lagrangian density $\cal L$ and 
$\Pi_i^\mu \equiv \frac{\delta \cal L}{\delta \partial_\mu \Phi^i}$. 
We shall call the $\Pi\delta\Phi$ part of the current the {\it 
rigid} part, as this is the only part that contributes for  
rigid internal symmetries.  
In the case of space-time symmetries we also have $V^\mu$, where  
$\delta {\cal L} = \partial_\mu V^\mu$ and the variation of 
the Lagrangian density is taken {\it without} using the Euler-Lagrange
equations for the fields. This situation will be familiar to the 
reader from the case of space-time translation symmetry where $V_\mu$ 
leads to the well-known term $-g_{\mu\nu}{\cal L}$ in the energy-momentum 
tensor\footnote{In the case of supersymmetry transformations we 
have to enlarge the space-time to a {\it super}-space with 
grassmanian coordinates
if we want to explicitly see those transformations as ``space-time''. 
Our approach is  based on standard space-time and 
we do not have an analog of $V^\mu = - {\cal L} \delta x^\mu$ therefore 
we derived the current by comparing the variation {\it off-shell} 
with the variation {\it on-shell} \cite{lop}.}.
For supersymmetric transformations the situation is more complicated 
simply because $V_\mu$ is more complicated and a major part of the task is to 
determine this quantity. Of course, $V_\mu$ is related to the second-last 
term in the superfield expansion, but knowing this does not appreciably 
simplify its computation. 
A second task is to write the full $J^\mu$ in terms of fields and 
their conjugate momenta\footnote{The difficulty of expressing 
Noether space-time currents,
and therefore {\it super}-currents, in terms of fields and their conjugate 
momenta was pointed out to me by Steven Weinberg.}. 
When $J^\mu$ is written in this way 
the transformations, the Hamiltonian and the central charge can be 
obtained from the canonical commutation relations. 
It is the result of these calculations that we wish to present 
in this letter.  

\noindent
For the Seiberg-Witten centre the relevant formula is

\be\label{q1q2}
\{ Q_{1 \alpha} \; , \; Q_{2 \beta} \} = 2 i \epsilon_{\alpha \beta} \; Z
\qquad
Q_{L \alpha} = \int d^3 x J_{L \alpha}^0 (x)
\quad L = 1 , 2
\ee
where  
$\epsilon_{1 2} = 1 = - \epsilon_{2 1}$ is understood,
the Lie brackets $\{ , \}$ stand for Poisson brackets 
and the charges $Q_{1\alpha}$ and $Q_{2\beta}$ are the generators of 
the two supersymmetries in $N=2$. It is actually necessary to compute 
only $Q_{1\alpha}$, because $Q_{2\alpha}$ can be obtained from it 
by an R-transformation i.e. by letting $\psi\leftrightarrow  -\lambda$ 
and $v_\mu \rightarrow -v_\mu$. The charges $Q_{L \alpha}$ 
come from the low-energy U(1) effective Lagrangian density which
in component fields, up to four fermions and second derivatives 
of the fields, is given by

\bea
{\cal L} & = &
{\rm Im}{\Big [}-{\cal F}'' (A)
   [ \partial_\mu A^\dagger \partial^\mu A
     + \frac{1}{4} v_{\mu \nu}v^{\mu \nu}
     + \frac{i}{8} v_{\mu \nu}v^{* \mu \nu}
     + i \psi  \not\!\partial {\bar \psi} 
     + i \lambda  \not\!\partial {\bar \lambda}
     - (F^\dagger F + \frac{1}{2} D^2)] \nonumber \\
&  & + {\cal F}''' (A) 
    [ \frac{1}{\sqrt 2} \lambda \sigma^{\mu \nu} \psi v_{\mu \nu}
      -\frac{1}{2} (F \psi^2 + F^\dagger \lambda^2)
      + \frac{i}{\sqrt 2} D \psi \lambda ] 
    + {\cal F}'''' (A) 
    [\frac{1}{4} \psi^2  \lambda^2] {\Big ]} 
\eea
where $A , \psi, F$ and $v_\mu , \lambda , D$ are the chiral and
vector N=1 multiplets respectively, we have chosen the temporal 
gauge $v_o=0$, fermions are in Weyl notation and  
${\cal F}(A)$ is the holomorphic 
prepotential which classically reduces to $\frac{1}{2} \tau A^2$, with
the complex coupling constant\footnote{Note that our $\tau$ 
is normalized to $\tau / 4\pi $ of Seiberg-Witten.
Note also that we keep $\theta \ne 0$ only as a computational tool
even if the physics is not affected by it.} 
$\tau = \tau_R + i \tau_I = \frac{\theta}{8 \pi^2} + \frac{i}{g^2}$.

\noindent
As we wish to compare the classical and effective central charges of 
the N=2 supersymmetry, we first re-obtain the classical result
by starting from the classical version of (4), namely

\be
{\cal L} = 
{\rm Im}
 {\Big [} - \tau [\partial_\mu A^\dagger \partial^\mu A
+ \frac{1}{4} v_{\mu \nu}v^{\mu \nu}
+ \frac{i}{8} v_{\mu \nu}v^{* \mu \nu} 
+ i \psi  \not\!\partial {\bar \psi} + i \lambda  \not\!\partial {\bar \lambda}
-  (F^\dagger F + \frac{1}{2} D^2)] {\Big ]}
\ee
By direct inspection we find for $Q_{1\alpha}$ supersymmetry 

\bea
\Pi_i^\mu \delta_1 \Phi^i &=&  \pi^\mu_A \delta_1 A  + 
\Pi^{\mu \nu} \delta_1 v_\nu +\delta_1 \bar\psi \pi^\mu_{\bar\psi} 
\label{nc1} \\
V_1^\mu &=&   \pi^\mu_A \delta_1 A
- \frac{i}{2} \tau^* \epsilon_1 \sigma_\nu \bar\lambda 
v^{* \mu \nu} \label{nc2}
\eea
where  we have written the conjugate momenta for each of the dynamical 
fields (in the temporal gauge for $v_\mu$). Thus for the vector field 
$\Pi^{\mu \nu} = \frac{i}{2} (\tau \hat{v}^{\mu \nu} - 
\tau^{*} \hat{v}^{\dagger \mu \nu})$, where $\hat{v}^{\mu \nu}$ and 
$\hat{v}^{\dagger \mu \nu}$ are the self-dual and antiself-dual projections
of $v^{\mu \nu}$ respectively,  
and for the fermions $\bar\psi$ and $\bar\lambda$ are regarded as 
the fields and $i \tau_I {\bar\sigma}^0 \psi$ and 
$i \tau_I {\bar\sigma}^0 \lambda$ as their conjugate momenta. 

\noindent
According to (2) the full supersymmetry current and charge then read

\be\label{j1c}
J_1^\mu = \Pi^{\mu \nu} \delta_1 v_\nu 
+ \delta_1 \bar\psi \pi^\mu_{\bar\psi} 
+ \frac{i}{2} \tau^* \epsilon_1 \sigma_\nu \bar\lambda v^{* \mu \nu}
\ee
and

\be
Q_{1 \alpha} = 
 \int d^3 x {\Big (}\Pi^i \delta_{1 \alpha} v_i
+ \delta_{1 \alpha} \bar\psi \pi_{\bar\psi} 
+ \frac{i}{2} \tau^* (\sigma_\nu \bar\lambda)_{\alpha} v^{* 0 i}{\Big )}
\ee
respectively. Note that in the total current the scalar terms 
$\pi_A\delta_1 A$ have canceled. 

\noindent
The charge $Q_{1\alpha}$ correctly generates the $\epsilon_1$ supersymmetry 
transformations of the fields:
for $v_i$ and $\bar\psi$ this is obvious; 
for $A$ we see that $\delta_1 \bar\psi$
contains $\pi_A$ and the commutation gives the right factors;
for $\lambda$ and $\psi$ one has to realize that 
$\delta_1 v_i$ contains $\bar\lambda$ and that we 
are looking for the transformations of the momenta 
$\pi_{\bar\lambda} = i \tau_I {\bar\sigma}^0 \lambda$ and 
$\pi_{\bar\psi} = i \tau_I {\bar\sigma}^0 \psi$. 
An important feature to notice is that the Noether charges  
automatically produce the {\it on-shell} transformations \cite{ior2}.

\noindent
By commuting the charge $Q_{1\alpha}$ with the charge $Q_{2\alpha}$ 
obtained by R-symmetry according to Eq. (\ref{q1q2}) we obtain, as might 
be expected, a total divergence, namely

\bea
\{ Q_{1 \alpha} \; , \; Q_{2 \beta} \} &=& 
- \int d^3 x \partial_i {\Big [}
(2 \sqrt{2} \Pi^i A^\dagger + \sqrt{2} v^{* 0 i} A^\dagger_D) 
\epsilon_{\alpha \beta}
+ \tau^* \epsilon^{0 i j k} (\sigma_j \bar{\lambda})_\alpha
(\sigma_k \bar{\psi})_\beta {\Big ]} \nonumber \\
&=& - \epsilon_{\alpha \beta} 2 \sqrt2  \int d^2 \vec{\Sigma}
\cdot (\vec{\Pi} A^\dagger + \vec{B} A^\dagger_D) \label{Zclass}
\eea
where $d^2\vec\Sigma$ is the measure on the sphere at infinity, 
$B^i = \frac{1}{2} \epsilon^{0 i j k} v_{j k}$ and 
$A^\dagger_D = \tau^* A^\dagger$ is the classical analogue of the {\it dual}
of the scalar field, which in Seiberg-Witten is 
$A^\dagger_D = {\cal F}^{\dagger '} (A^\dagger)$. We have 
made the usual assumption that $\bar\psi$ and $\bar\lambda$ fall off 
at least like $r^{-{3\over 2}}$ and have implemented the Gauss law 
as an identity. 

\noindent
For the effective Lagrangian (4) much more labor is needed to write the
variation of $\cal L$ as a pure divergence\cite{ior2}. At the end of 
the computation the {\it rigid} current and $V^\mu$ turn out to be

\bea
\Pi_i^\mu \delta_1 \Phi^i &=&  \pi^\mu_A \delta_1 A  + 
\Pi^{\mu \nu} \delta_1 v_\nu +\delta_1 \bar\psi \pi^\mu_{\bar\psi} \\
V_1^\mu &=&  \pi^\mu_A \delta_1 A 
- \frac{i}{2} {\cal F}^{'' \dagger} \epsilon_1 \sigma_\nu \bar\lambda 
v^{* \mu \nu} 
+ \frac{1}{2\sqrt2}  {\cal F}^{''' \dagger}
\epsilon_1 \sigma^\mu \bar\psi \bar\lambda^2
\eea
Here again we have written the conjugate momenta of the fields. We 
notice that the {\it rigid} current is formally the same 
as in the classical case and that $V_1^\mu$ differs only in the last 
term (containing $\bar\psi\bar\lambda^2$). Thus these currents    
correctly reduce to the ones in (\ref{nc1}) and (\ref{nc2})
when the classical limit
is taken: ${\cal F}(A) \rightarrow \frac{1}{2} \tau A^2$. However, 
the formal resemblance 
masks the fact that the momenta and ${\cal 
F''}^\dagger$ are quite different when expressed as explicit 
functions of the fields and their derivatives. 
In particular the conjugate momentum of $v_\mu$ has a much more complicated 
expression involving all the fermions:

\noindent
$
\Pi^{\mu \nu} = \frac{i}{2}({\cal F}'' {\hat v}^{\mu \nu} - 
     {{\cal F}''}^{\dagger} {\hat v}^{\dagger \mu \nu})
- \frac{i}{\sqrt2}({\cal F}''' \lambda \sigma^{\mu \nu} \psi 
    -{{\cal F}'''}^{\dagger} \bar\lambda \bar\sigma^{\mu \nu} \bar\psi) .
$

\noindent
According to (2) the full current is then

\be
J_1^\mu = \Pi^{\mu \nu} \delta_1 v_\nu 
+ \delta_1 \bar\psi \pi^\mu_{\bar\psi} 
+ \frac{i}{2} {\cal F}^{'' \dagger} 
\epsilon_1 \sigma_\nu \bar\lambda v^{* \mu \nu}
- \frac{1}{2\sqrt2}  {\cal F}^{''' \dagger}
\epsilon_1 \sigma^\mu \bar\psi \bar\lambda^2
\ee
where once again the scalar contributions $\pi_A\delta_1 A$ have 
canceled.  The charge is 

\be
Q_{1 \alpha} = 
 \int d^3 x {\Big(} \Pi^i \delta_{1 \alpha} v_i
+ \delta_{1 \alpha} \bar\psi \pi_{\bar\psi} 
+ \frac{i}{2}  {\cal F}^{'' \dagger} 
(\sigma_\nu \bar\lambda)_{\alpha} v^{* 0 i}
- \frac{1}{2\sqrt2}  {\cal F}^{''' \dagger}
(\sigma^0 \bar\psi)_{\alpha} \bar\lambda^2{\Big )}
\ee
Again this charge reproduces the right transformations by simply applying
the same procedure outlined in the classical case and carefully handling 
the cubic fermion terms, which recombine to give the 
{\it on-shell} dummy fields \cite{ior2}.

\noindent
According to Eq. (\ref{q1q2}) we commute the charge $Q_{1\alpha}$ with the 
charge $Q_{2\alpha}$  obtained by R-symmetry, and find that 

\bea
\{ Q_{1 \alpha} \; , \; Q_{2 \beta} \} &=& 
- \int d^3 x \partial_i {\Big [}
(2 \sqrt{2} \Pi^i A^\dagger + \sqrt{2} v^{* 0 i} A^\dagger_D) 
\epsilon_{\alpha \beta}
+ {\cal F}^{\dagger ''} \epsilon^{0 i j k} (\sigma_j \bar{\lambda})_\alpha
(\sigma_k \bar{\psi})_\beta {\Big ]} \nonumber \\
&=& - \epsilon_{\alpha \beta} 2 \sqrt2  \int d^2 \vec{\Sigma}
\cdot (\vec{\Pi} A^\dagger + \vec{B} A^\dagger_D) \label{Zeff}
\eea
where again we have made the usual assumption that the fermion 
fields drop off at least like $r^{-{3\over 2}}$ and have implemented 
the Gauss law as an identity.

\noindent
Note that the two expressions (\ref{Zclass}) and (\ref{Zeff})
have the same {\it form} and they only differ by replacing classical 
fields and momenta (and their dual) by their quantum counter-parts. 
Of course this does not mean that the two centres are equal (as is
expected to be
the case for N=4 supersymmetry where the beta function is identically zero)
but only that supersymmetry has protected the 
classical {\it form} and therefore the BPS mass-formula equally applies 
to the quantum case. 

\noindent
Finally one has to evaluate (\ref{Zclass}) and (\ref{Zeff}) 
explicitly in the limit $A^\dagger\rightarrow 
a^*$ and $A_D^\dagger\rightarrow a^*_D$ for $r \rightarrow \infty$ 
where $a^*$ and $a_D^*$ are c-numbers, and are non-zero in the 
spontaneously broken case. Up to an irrelevant numerical scale-factor 
\footnote{$Z \rightarrow  (i \sqrt2 \alpha)^{-1} Z$ 
and $\alpha$ depends on the 
conventions one uses to compute the integrals.} we obtain  

\be
Z = n_e a^* + n_m a^*_D
\ee
in agreement with (1). 
 
\noindent
{\bf Acknowledgments.}

\noindent
This work was partially supported by FORBAIRT (Ireland) and by Istituto 
Italiano per gli Studi Filosofici (Naples, Italy).
The author would 
like to thank L. O'Raifeartaigh, I. Sachs and M. Magro for valuable 
help and suggestions.

\end{document}